\documentclass[aps,prl,twocolumn,showpacs,showkeys,groupedaddress,amsmath,amssymb]{revtex4}
\usepackage{graphicx}
\usepackage{color}
\usepackage{bm}		% bold math
\usepackage{ulem}
\usepackage{amsmath}
\usepackage{physics}
\begin{document}

\title{Plaquette Valence Bond Theory of High-Temperature Superconductivity}

\author{M. Harland}
\affiliation{Institut f\"ur Theoretsche Physik, Universit\"at
Hamburg, Jungiusstra{\ss}e 9, 20355, Hamburg, Germany}

\author{M.~I. Katsnelson}
\affiliation{Radboud University Nijmegen, Institute for Molecules and Materials,
6525AJ, Nijmegen, the Netherlands}

\author{A.~I. Lichtenstein}
\affiliation{The Hamburg Centre for Ultrafast Imaging, Luruper Chaussee 149, Hamburg 22761, Germany}
\affiliation{Institut f\"ur Theoretsche Physik, Universit\"at
Hamburg, Jungiusstra{\ss}e 9, 20355, Hamburg, Germany}

%\date{\today}

\begin{abstract}
We present a strong-coupling approach to the theory of high-temperature superconductivity based on the observation of a quantum critical point in the plaquette within the $t,t'$ Hubbard model. The crossing of ground state energies in the $N=2,3,4$ sectors occurs for parameters close to the optimal doping. The theory predicts the maximum of the d$_{x^2-y^2}$-wave order parameter at the border between localized and itinerant electron behavior and gives a natural explanation for the pseudo-gap formation via soft-fermion mode related to local singlet states of the plaquette in the environment. Our approach follows the general line of resonating valence bond theory stressing a crucial role of singlets in the physics of high-T$_c$  superconductors, but focuses on the formation of {\it local} singlets, similar to phenomena observed in {\it frustrated} one-dimensional quantum spin models.
\end{abstract}

\maketitle

%\section{Introduction}
%{\it Introduction}
After 30 years history of extremely intensive experimental \cite{Bednorz, Damascelly,Orenstein,Davis} and theoretical  
\cite{PWA, PWAbook, Scalapino, Pines, ImadaRMP,Dagotto} studies of the high-temperature superconductivity (HTSC) in
copper oxides we are still far from understanding the basic mechanism of this fascinating phenomenon.
Taking into account the enormous number of researchers involved in this field, one can assume that almost all
possible ideas were expressed and that the main problem is just to select the basic simple concepts from the
pile of available theoretical results. The most ambitious attempt was made by P. W. Anderson
who emphasized with his RVB (resonating valence bond) theory the crucial importance of strong electron correlations, the tendency to singlet spin state formation and the non-Fermi liquid character of the normal phase \cite{PWAbook}. Unfortunately, details of his original approach, such as suppression of interlayer hopping in the normal phase as the main factor of superconductivity, seems to contradict experimental data \cite{vanderMarel}.
The latest version of the RVB theory is presented in Ref. \cite{PWA2}.
We believe, that the main assumption of the strongly correlated limit as the base of understanding
the high-temperature superconductivity is correct, as well as emphasizing a crucial role of spin singlet states, but important details were missing.
Below we will present arguments for the thesis, that the minimal object of HTSC-theory is the plaquette in the so-called
effective $t,t'$ Hubbard model \cite{OKA}, rather than the conventional atomic limit typical for the theory
of Mott insulators  \cite{PWAbook, ImadaRMP}.  The best practical realization of this atomic based theory is
the dynamical mean-field theory (DMFT) \cite{DMFT}.
The obvious minimal generalization in the case of d$_{x^2-y^2}$-wave pairing is a cluster DMFT (CDMFT)
scheme \cite{CDMFT,JarrellRMP}.

Since the first plaquette CDMFT calculation of d$_{x^2-y^2}$-wave superconducting
order together with antiferromagnetic fluctuations \cite{CDMFT}, there have been many
calculations for different cluster sizes and geometries based on
continuous-time Quantum Monte Carlo (CTQMC) or exact diagonalization (ED) solvers \cite{Kotliar,Ferrero2, Ferrero, Gull, Jarrell2,Gull2,Tremblay, JarrellRMP,Maier,Potthoff,Millis,Civelli,Kancharla,Okamoto,Tremblay4,Tremblay3,ChenGull,Tremblay2}.
Unfortunately, the basic qualitative feature of the many-body states in the plaquette were hidden in
computational details. The main aim of this work is to present a simple and transparent strong coupling theory of the d$_{x^2-y^2}$-wave superconductivity (a {\it minimal} consistent many-body model) based on the plaquette energy spectrum peculiarity, namely the ``quantum critical point'', that merges two singlets and two doublets. These states of the doped plaquette are different from those
discussed in the resonating valence-bond theory \cite{PWAbook,PWA2}. The main point is that the quantum critical point discussed here is related to the
formation of {\it local} valence bonds in the {\it frustrated} quantum spin model \cite{Affleck}. Therefore, the optimal
superconducting states are located on the border between localized and delocalized (resonating) plaquette valence bonds.
Here we follow a bottom-up approach starting with isolated plaquette and building stepwise a more complicated environment.

An important theoretical problem is to find a minimal and generic electronic-structure model of
cuprate superconductors. From band-structure calculations \cite{OKA,Pavarini}  we can
quite safely reduce it to an effective one-band model with long-range hopping.
We use a standard parametrization  of the tight-binding model for YBa$_{2}$Cu$_3$O$_7$\cite{OKA,Pavarini}
with the next-nearest neighbor hopping: $t'/t=-0.3$ and $t$ as unit of our energies.
The local Hubbard interaction parameter $U$ is of the order of the band-width $W=8t$.
The corresponding $t, t'$ Hubbard model on the square lattice reads
\begin{equation}
H=-\sum_{ij}t_{ij}c_{i\sigma }^{\dag}c_{j\sigma }+\sum_{i}U n_{i\uparrow
}n_{i\downarrow }
\end{equation}
where $t_{ij}$ is an effective hopping and $U$ the local Coulomb
interaction. The operators $c_{i\sigma }^{\dag}$, $c_{i\sigma }$ create and annihilate fermions at site $i$
with spin $\sigma=\uparrow (+),\downarrow (-)$, respectively and the occupation operator is $n_{i\sigma}=c_{i\sigma }^{\dag}c_{i\sigma}$.

%
%\section{Isolated Plaquette}
{\it Isolated Plaquette} -- We start the discussion with electronic states in the isolated Hubbard plaquette.
The optimal doping for high-temperature superconductivity is of the order of $15\%$ of
holes per site for almost all cuprate materials. This gives us an average number of electrons
per site of 0.85, which corresponds to 3.4 fermions per 4-site plaquette in the crystal.
We argue, that this is related to 3-electron states of the isolated plaquette, since
particle-hole asymmetry introduced by the next-nearest neighbors hopping $t'$,
with moderate values of $U$ and certain fixed chemical potentials ($\mu$) result in an occupation per plaquette
of a crystal, that is very close to the optimal value of 3.4 electrons.

\begin{figure}[htbp]
\includegraphics[angle=0,width=1.0\columnwidth]{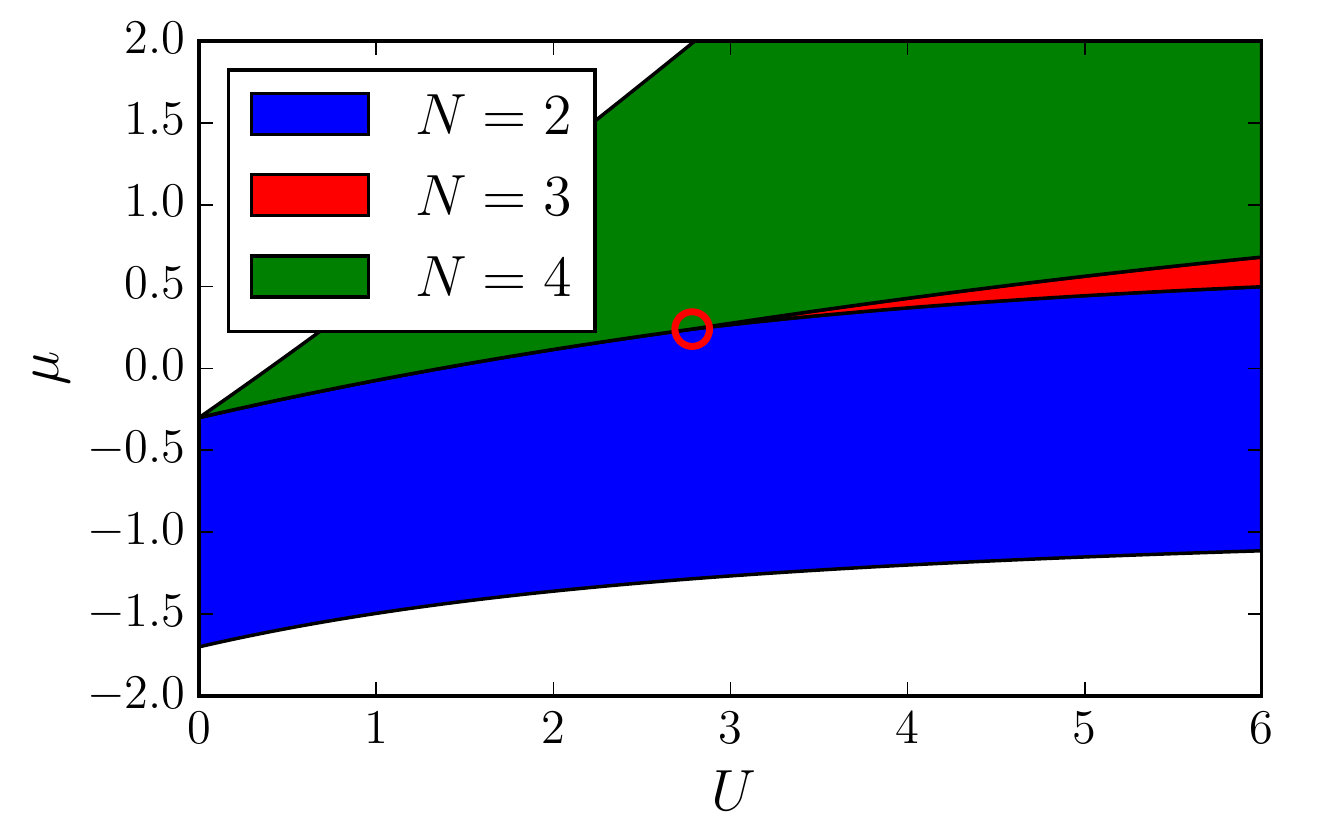}
\caption{(Color online) Zero temperature phase diagram for isolated plaquette as function of Hubbard U
and chemical potential $\mu$ close to 3-particle occupation and $t'/t=-0.3$.}
\label{phase_plaquette4}
\end{figure}

The Hamiltonian of the isolated plaquette reads
\begin{equation}
H_p=-\sum_{(i,j)=1..4} h^0_{ij}c_{i\sigma }^{\dag}c_{j\sigma }+\sum_{i=1..4}U n_{i\uparrow
}n_{i\downarrow },
\end{equation}
\begin{equation}
\hat{h}_0  =\left(
\begin{array}{cccc}
\mu &t &t' & t\\
t  & \mu& t & t' \\
t' & t  & \mu &t  \\
t  & t' & t  & \mu
\end{array}
\right),
\label{tplaquete}
\end{equation}
where we included chemical potential in the diagonal part of $h^0_{ij}$.
The energy spectrum of the isolated plaquette near the 3-electron filling is quite unusual.
We present in Fig.~\ref{phase_plaquette4}
regions in the $U-\mu$ space, where the ground state corresponds to three plus-minus one electron.
The one-electron spectrum consists of the following four states with the energies:
$\pm 2t - t'- \mu$ and double-degenerate $t'- \mu$. At zero interaction $U=0$,
there is no stable ground state with three electrons, in a sense that one can add or remove
one electron without a change of the thermodynamic potential.
Starting from some critical interaction strength $U \approx 3$
there is a small region (red part of the Fig.~\ref{phase_plaquette4}), where the plaquette
ground state corresponds to three electrons separated by energy gaps from the states
with $N=2$ and $N=4$.
Importantly, this $N=3$ ground state is fourfold degenerate corresponding
to two doublets in the sectors $(2_\uparrow,1_\downarrow)$ and $(1_\uparrow,2_\downarrow)$,
which we will call $\ket{X}$ and $\ket{Y}$  states according to their symmetry.
Moreover there is a critical point (red circle in Fig.~\ref{phase_plaquette4}) where all
three sectors with 2, 3, and 4 electrons have the same ground state energy and
form sixfold degenerate ground-state multiplet consisting of two singlets of the sectors $(1_\uparrow,1_\downarrow)$
and $(2_\uparrow,2_\downarrow)$ together with two doublets of the three-electron sectors. For standard values of $t'/t=-0.3$ this critical point has the
coordinates ($U=2.78$, $\mu=0.24$).
We think that this critical point of the plaquette has crucial importance for the physics of the
strong-coupling  d$_{x^2-y^2}$ wave superconductivity. The importance of these three many-body
states of the plaquette CDMFT was first discussed for the $t-J$ model \cite{Kotliar}. In that case there is an additional triplet state in the $N=4$
sector, which appeared in our case only for $U/t \geq 6$. In the valence bond DMFT approach  \cite{Ferrero2} a
similar crossing of different many-body states  appears in a correlated dimer. The idea of a quantum critical
point was also discussed in Refs. \cite{Jarrell2,Tremblay4}. Here we will demonstrate, via bottom-up approach, that this is {\it the} key ingredient of a consistent minimal picture of HTSC.

If we approach this critical point from the red region with the $N=3$ ground state, then
the one electron density of states (DOS) at the Fermi energy diverges for both, electron and hole sides,
due to transitions from the fourfold-degenerate $N=3$ ground state to singlets of $N=2$ (hole side)
and $N=4$ (electron side) with zero excitation energy. The corresponding spectral weights
(normalization of the $\delta$-functions) are equal to 0.26 and 0.13 for hole and electron sides,
respectively which already introduces an important electron-hole asymmetry.  We will see below that this plaquette quantum critical point results in a formation of a ``soft''-fermion mode (i.e. a DOS peak at the Fermi energy), when passing
to a crystal made of plaquettes. We argue, that these soft-fermions favor the formation of the
d$_{x^2-y^2}$-wave superconducting pairing at low temperatures
and of the pseudogap at high-temperatures. At smaller $t'$ this critical point shifts to larger $U$ and
at  $t'/t=0$ it is equal to ($U=4.58$, $\mu=0.72$). It is worthwhile to point out that at optimal
values of  $t'$ antiferromagnetic order is suppressed due to frustrations.
As soon as we add a fermionic bath to the plaquette within the spirit of CDMFT or
density matrix embedding theory (DMET) \cite{DMET}   with only four bath sites,
a stable singlet solution is formed with almost equal mixture of all
$N=2, 3, 4$ sectors, which again is favorable for the superconducting state as will be shown below.
%%%%%%%%%%%%%%  Fig1 %%%%%%%%%%%%%

\begin{figure}[htbp]
\includegraphics[angle=0,width=0.6\columnwidth]{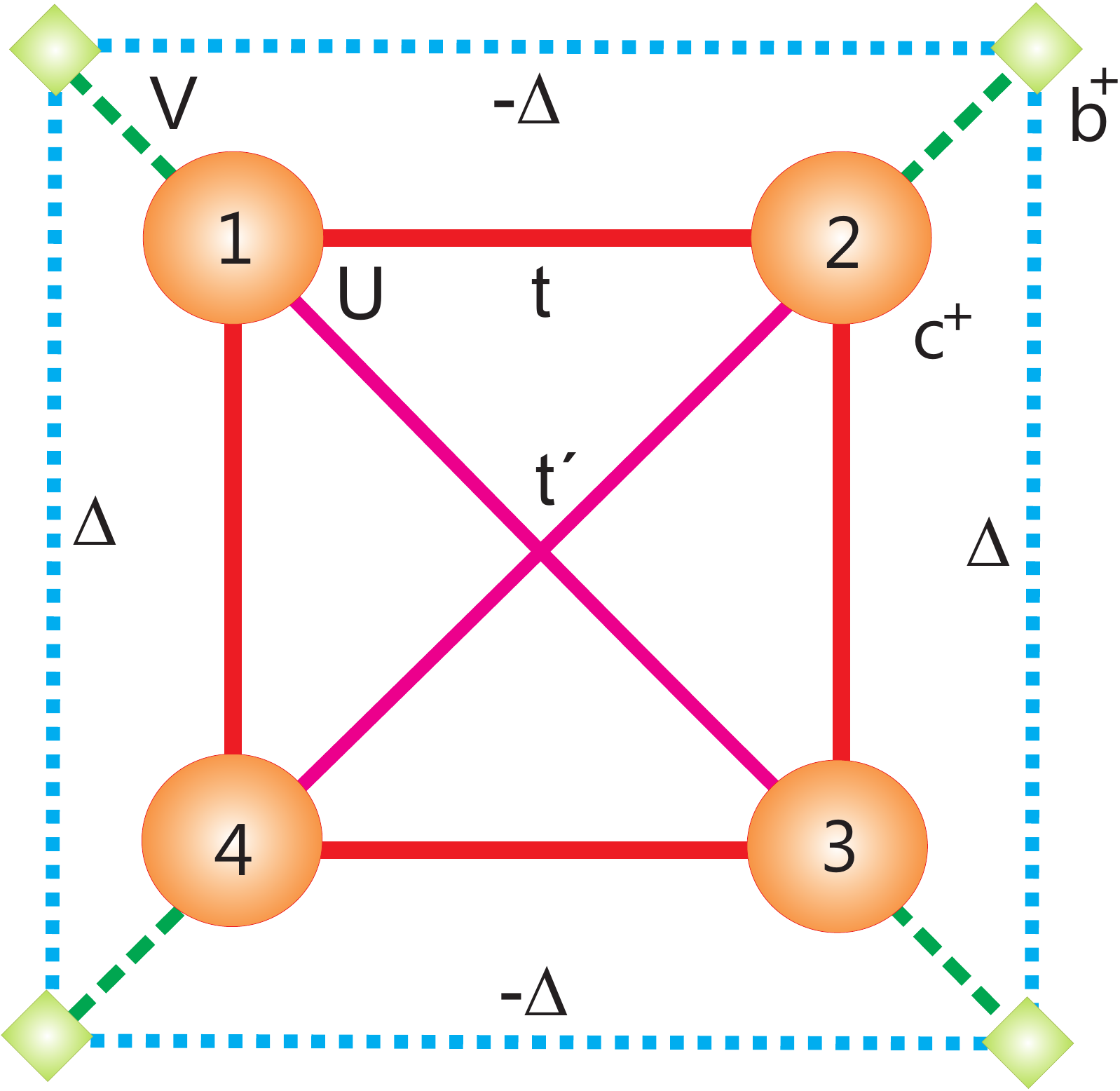}
\caption{(Color online) Sketch of the plaquette in the four-site bath with superconducting d$_{x^2-y^2}$-wave order parameter.}
\label{plaquette4bath}
\end{figure}

%%%%%%%%%%%%%%%%%%%%%%%%%%%%%%%
%%%%%%%%%%%%%%  DMET: Plaquette in bath %%%%%%%%%%%%%
%\section{Plaquette in Bath}
{\it Plaquette in Bath} -- The appearance of the DOS peak at the Fermi energy leads to a universal instability in a sense,
that the susceptibility diverges in many different channels (magnetic, superconducting,
charge density wave, etc.). From the weak-coupling side this was discussed in the framework
of the van Hove scenario of HTSC \cite{IKK,Kampf,Metzner}. To study the interplay of different
instability channels from the strong-coupling perspective we introduce a simple model of an
embedded plaquette in the spirit of DMET \cite{DMET}. To this aim we add to the plaquette
Hamiltonian a hybridization with four fermionic bath states (i.e. one bath state per corner of the plaquette,
see Fig.~\ref{plaquette4bath}) and use the exact diagonalization technique
(Lanczos scheme with a Hilbert space size of $2^{16}$, without any symmetry restrictions) with
different symmetry breaking fields acting on the bath fermions $b_{i\sigma }^{\dag}, b_{j\sigma }$,
namely d-wave pairing, singlet magnetic states on the bonds,
and the conventional Neel antiferromagnetic state:
\begin{equation}
h_d =\sum_{\sigma=\uparrow,\downarrow,  i=1..4 } (-1)^i \sigma \Delta_d  (b_{i,\sigma }b_{i+1,-\sigma} + h.c.),
\label{dsc}
\end{equation}
\begin{equation}
h_s =\sum_{\sigma=\uparrow,\downarrow,  i=1..4  } (-1)^i \sigma \Delta_s  (b_{i,\sigma }^{\dag}b_{i+1,-\sigma} + h.c.),
\label{mbs}
\end{equation}
\begin{equation}
h_m =\frac{1}{2} \sum_{\sigma=\uparrow,\downarrow,  i=1..4  } (-1)^i \sigma \Delta_m  b_{i,\sigma }^{\dag}b_{i,\sigma}.
\label{afm}
\end{equation}
Here we assume periodic boundary conditions, i.e. for $i=4$ we define $i+1=1$. Simultaneously we switch on small fields
$\Delta_d=\Delta_s=\Delta_m=0.01 t$ and calculate numerically the corresponding susceptibilities of the plaquette.
The hybridization $V$ between the fermions $c_{i\sigma }^{\dag}$ and $b_{j\sigma }^{\dag}$ breaks the sixfold
degeneracy of the plaquette's quantum critical point and without external fields it results in a singlet ground state
(see Fig.~\ref{plaquette4bath}).
The d$_{x^2-y^2}$-wave superconducting (Eq.~(\ref{dsc})) and magnetic bond-singlet (Eq.~(\ref{mbs})) external fields respect quantum entanglement of the singlet character of the ground state, whereas the Neel field Eq.~(\ref{afm}) destroys it. Being classical in its nature the Neel state is
expected to be most stable for sufficiently strong coupling with the environment $V$ \cite{HH_dbethe} or high temperatures \cite{Otsuki}.
%%%%%%%%%%%%%%  Fig2 %%%%%%%%%%%%%%
%%%%%%%%%%%%%%%%%%%%%%%%%%%%%%%
%%%%%%%%%%%%%%  Fig3 %%%%%%%%%%%%%%
\begin{figure}[htbp]
\includegraphics[angle=0,width=0.9\columnwidth]{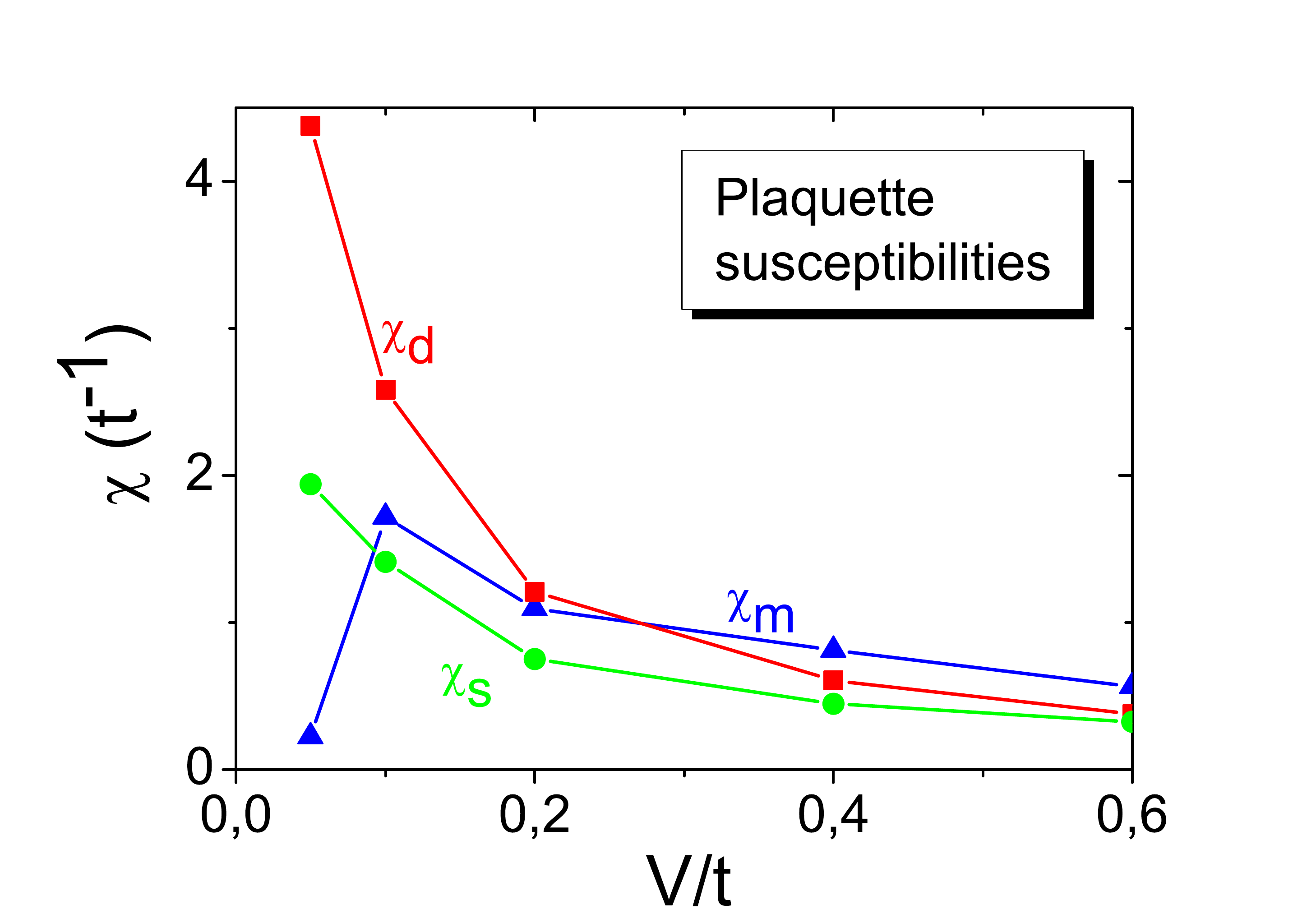}
\caption{(Color online) Superconducting d$_{x^2-y^2}$-wave ($\chi_d$), singlet bond order ($\chi_s$) and antiferromagnetic ($\chi_m$)
susceptibility of the plaquette in a bath as a function of the hybridization $V$ for $U=3$ and $\mu=0.27$.}
\label{plaquette-susc}
\end{figure}
%%%%%%%%%%%%%%%%%%%%%%%%%%%%%%%
For an infinite system different types of order can be found by studying the divergence of the corresponding
susceptibility. Since in DMET we deal with finite systems, the susceptibilities remain finite up to
zero temperature and we will assume, that the largest susceptibility of the cluster, shown in Fig.~\ref{plaquette4bath},
signals the corresponding order of the crystal. The computational results are shown in Fig.~\ref{plaquette-susc}
as function of the hybridization parameter $V$. One can see that d$_{x^2-y^2}$-wave superconducting pairing
always wins in comparison with the singlet bond pairing and is more favorable than the Neel order for
$V \le 0.2$. The self-consistent plaquette-Bethe DMFT for the cluster case corresponds to  $V = 0.1$, thus
the singlet ground state near the plaquette's quantum critical point favors d$_{x^2-y^2}$-wave superconductivity
rather than magnetic ordering. This result agrees well with a large scale CDMFT calculations
for optimal doping \cite{Civelli,Kancharla}.

\begin{figure}[htbp]
\includegraphics[angle=0,width=0.7\columnwidth]{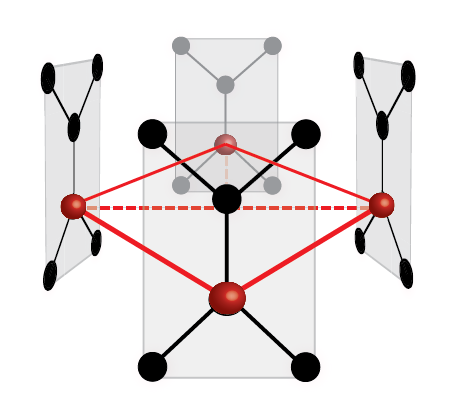}
\caption{(Color online) Sketch of a plaquette-Bethe lattice with connectivity q=2. Only one plaquette is shown for simplicity.}
\label{plaquette-bethe}
\end{figure}

%%%%%%%%%%%%%%%%%%%%%%%%%%%%%%%
%%%%%%%%%%%%%%  Plaquette-Bethe %%%%%%%%%%%%%
%\section{Plaquette-Bethe Lattice}
{\it Plaquette-Bethe Lattice} -- As a next step towards a more realistic description of the cuprate crystal, we consider
a plaquette-Bethe model, where all sites are arranged in quadrupole Bethe ``planes'' and interconnected
in a plaquette-like manner (see Fig.~\ref{plaquette-bethe}).
The plaquette CDMFT becomes exact for this mode, when the connectivity of the Bethe sublattice
$q$ tends to infinity. We obtain similar results as for the
double Bethe model for a two-site cluster \cite{Ruckenstein,HH_dbethe}.
The bath Green's function in this model reads
\begin{equation}
{\cal {\hat G}}^{-1}\left( i\omega \right) =\left( i\omega+\mu-{\hat h}_0 \right)^{-1} -t^2_b{\hat G}\left( i\omega \right) ,
\label{pbethe}
\end{equation}
where ${\hat G}\left( i\omega \right)$ is a $8 \times 8$ matrix of the superconducting Green's function for the plaquette with opposite spin-Nambu spinors in the bath.
We discretize the bath Greens's function with only four states similar to the DMET approach using the Lanczos scheme in order to
find the matrix Green's function for the superconducting states \cite{Kancharla,Civelli}.
%%%%%%%%%%%%%%  Fig4 %%%%%%%%%%%%%%
%%%%%%%%%%%%%%%%%%%%%%%%%%%%%%%
%%%%%%%%%%%%%%  Fig5 %%%%%%%%%%%%%%
\begin{figure}[htbp]
\includegraphics[angle=0,width=1.0\columnwidth]{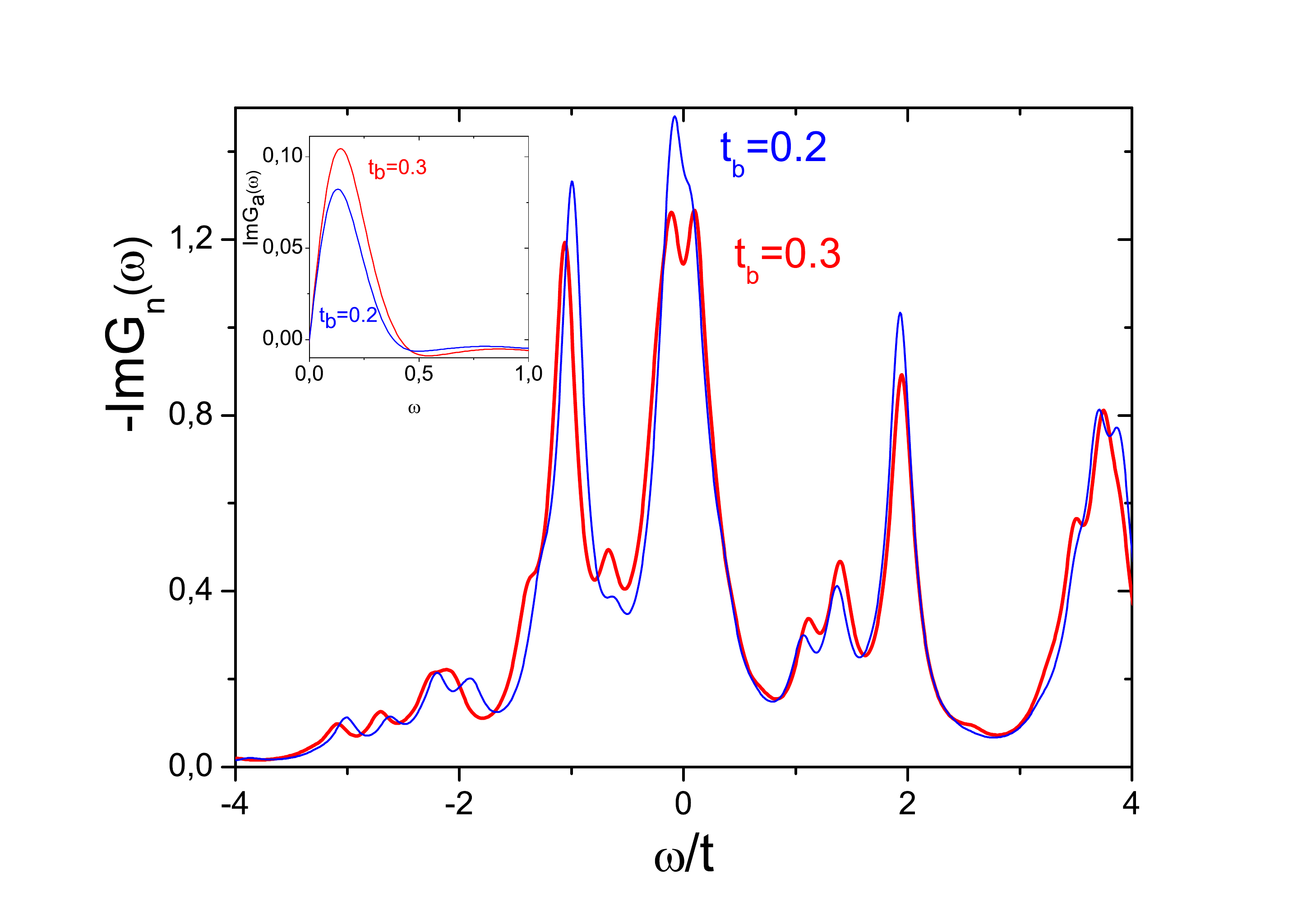}
\caption{(Color online) Local part of normal Green function for the d$_{x^2-y^2}$-wave solution in the plaquette Bethe lattice near optimal values of $t_b$ for $U=3$ and $n=0.85$. Inset: non-local part of anomalous Green function.}
%\caption{(Color online) Local imaginary part and non-local superconducting real part of the Green function for the d$_{x^2-y^2}$-wave solution in %the plaquette Bethe lattice for optimal values of $t_b$.}
\label{plaquette-green}
\end{figure}
%%%%%%%%%%%%%%%%%%%%%%%%%%%%%%%
As mentioned above, there is a sixfold degenerate ground state for $t_b=0$ at the quantum critical point.
At sufficiently small hybridizations, which correspond to small $t_b$ in the plaquette-Bethe model, the system
becomes metallic with a slightly broadened peak in the DOS at the Fermi energy.
The ground state is a d$_{x^2-y^2}$-wave superconducting at low temperatures\cite{Kancharla,Civelli}. However, when $t_b$ increases
a quantum phase transition occurs with the destruction of the singlet states and a formation of
the energy gap in the single-electron excitation spectrum. The latter can be observed in the normal
 part of the one-electron Green's function at $t_b=0.3$ (see Fig.~\ref{plaquette-green}).
The energy gap of the states can be estimated to 0.2$t$, that is an order of magnitude larger than the
superconducting gap. For $t=0.25$ eV \cite{OKA} this corresponds to a pseudogap of the order of 50 meV
which is observed experimentally  \cite{EXPpgap}.
In regard to the double-Bethe model this corresponds to a transition
of a quantum entangled singlet state to a classical Neel state  \cite{HH_dbethe}.
Importantly, the anomalous (superconducting) part of the Green's function
has a maximum exactly at this transition (see right inset in Fig.~\ref{plaquette-green}) .
%(see Fig.~\ref{plaquette-green}).
Similar behavior has been observed recently experimentally for the maximum of the superconducting order parameter at the localized - delocalized
transition point in the strongly correlated organic superconductors \cite{EXPgap} as well as for the BCS-BEC crossover in
cold-atom systems \cite{BCS_BEC}.

%%%%%%%%%%%%%%  CDMFT %%%%%%%%%%%%%
%\section{Plaquette CDMFT}
{\it Plaquette CDMFT} -- Finally, we perform standard CDMFT calculations using a strong-coupling continuous-time Quantum Monte Carlo solver \cite{CTQMC,TRIQS} in the normal state. Since a transition to the periodic plaquette in the crystal increases the
bandwidth by a factor of two due to doubling of the coordination numbers compared to the isolated plaquette, we increase the values of $U$ and $\mu$
by the factor of two. The natural energy unit is the bandwidth $W$ rather than the hopping $t$. Furthermore we use
the value of $U/t=6$ which approximately correspond to real cuprate materials \cite{expU}.

\begin{figure}[htbp]
\includegraphics[angle=0,width=1.0\columnwidth]{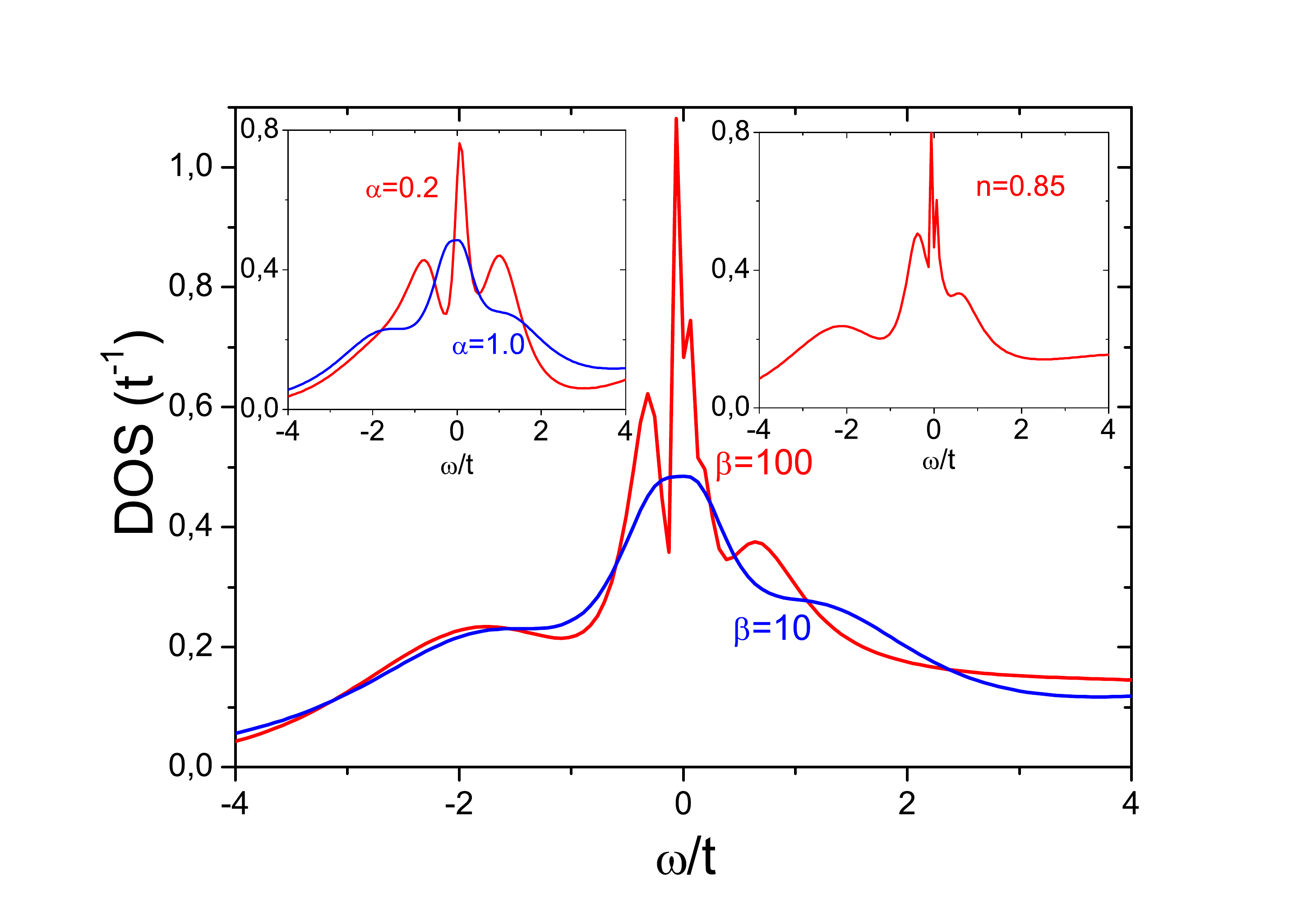}
\caption{(Color online) Density of states of the plaquette CDMFT for $U=6$ and $\mu=0.54$ for different temperatures.
Left inset: the different plaquette-lattice hoppings scaled by a factor $\alpha$ for $\beta =10$; Right inset: optimal doping $n=0.85$
for $\beta =100$.}
\label{htsc_plaquette-bethe}
\end{figure}

The calculated local DOS obtained by maximum-entropy analytic continuation \cite{MAXENT} is shown in Fig.~(\ref{htsc_plaquette-bethe}).
We observe, that for sufficiently high temperatures there is a broad peak at the Fermi energy originating from the plaquette quantum critical point. This relation is illustrated in the inset of Fig.~(\ref{htsc_plaquette-bethe}) where we artificially scaled the hopping
between plaquettes by a factor $\alpha$ ranging from 0 to the physical value 1.
%%%%%%%%%%%%%%  Fig2 %%%%%%%%%%%%%%
%%%%%%%%%%%%%%%%%%%%%%%%%%%%%%%

At lower temperatures a pseudogap is formed instead. The pseudogap is well known in all HTSC
materials and is considered as one of its most striking features \cite{EXPpgap}. Sometimes this pseudogap
is also considered to be the precursor of the superconducting gap (formation of incoherent Cooper pairs above $T_c$)
or as a smeared antiferromagnetic gap (shadow bands) \cite{EXPpgap}. However, both interpretations have
problems, when they get compared to experiments \cite{EXPpgap}.
Within our scheme it is natural to interpret this pseudogap as a pseudo-hybridization-gap similar to that
arising in Kondo lattices \cite{Bulla_rmp} or intermediate valence semiconductors \cite{Irkin_Kats_JETP86}.
From this point of view the pseudogap in HTSC materials originates from the Fano antiresonance due to embedding
of the soft-fermion mode of the plaquette (discussed above) into a continuous band spectrum of the lattice.
The role of soft-fermion modes (the hidden fermion) was discussed in Ref. \cite{Imada_HF}, however it was done without
any relation to the quantum critical point of the plaquette. This relation is the main message of our paper.

The density of states for optimal doping $n=0.85$ in self-consistent CDMFT calculations
corresponds to $\mu=1.2$ (see right inset of Fig.~\ref{plaquette-green}). It is very similar to that of
the fixed $\mu$ which corresponds to the quantum critical point of the isolated plaquette.
Furthermore, we calculated a low-temperature superconducting state with CDMFT, Lanczos solver and 8-bath sites and also with a CTQMC cluster solver
and found similar results to many other calculations
\cite{Kotliar,Ferrero2, Ferrero, Gull, Jarrell2,Gull2,Tremblay, JarrellRMP,Maier,Potthoff,Millis,Civelli,Kancharla,Okamoto,Tremblay4,Tremblay3,ChenGull,Tremblay2}.
One of the first analysis of many-body sector statistic for the plaquette cluster in a bath within
CTQMC calculations  \cite{Kotliar, Millis} and important interpretation of the pseudogap state  \cite{Tremblay3} should be also mentioned here.

%%%%%%%%%%%%%%  Conclusion %%%%%%%%%%%%%
%{\it Conclusions}
To conclude, we developed a picture of HTSC based on the existence of a quantum critical point at the
crossing of the ground state energies in the $N=2,3,4$ sectors within the plaquette for parameters close
to the optimal doping, $t'$ being of crucial importance.
Contrary to the original resonating valence bond (RVB) theory of Anderson  \cite{PWAbook},
we start with the $local$ valence bond formation in the doped plaquette.
The difference can be illustrated by comparison with the exactly solvable one-dimensional quantum spin model  \cite{Affleck}.
The prototype state for the RVB is the Bethe ansatz solution for the antiferromagnetic $S=1/2$ Heisenberg model
in the nearest-neighbor approximation. For the model with first- and second-neighbor interactions,
$J_2/J_1=1/2$, the ground state is known exactly, too and it can be represented as a product state of
local valence bond singlets \cite{Affleck}. The second-nearest hopping $t'$ seems to play a similar role
in the Hubbard model. The optimal superconducting region is related to a localized-delocalized transition of
plaquette valence bond states in the plaquette. It would be interesting to describe the formation of global singlet states with
plaquette valence bond states using the matrix product scheme  \cite{MPS}, since the CDMFT scheme
break translational symmetry.

Formation of the soft-fermion mode near the optimal doping has an analog in the weak coupling
theory within the van Hove scenario of HTSC \cite{IKK_PRL}. Due to the formation of flat bands of many-body
origin \cite{IKK_PRL,Yudin_PRL} there is a whole region of parameters $t', U, \mu$ where the Fermi-liquid
description is broken. However, we believe that the strong-coupling description presented here is more
relevant for real HTSC materials, which are characterized by quite large values of $U$ \cite{expU}.

%%%%%%%%%%%%%%  Acknowledgment %%%%%%%%%%%%%
We thank  Boris Altshuler, Ole Andersen, Phil Anderson, Antoine Georges, Andy Millis, Michael Potthoff, Tim Wehling, and Jens Wiebe
for helpful discussions as well as
Tim Berberich and Maria Valentyuk for assistance with the figures, Lewin Boehnke for the MaxEnt code and
Satoshi Okamoto for Lanczos SC code.
 A.~I.~L. acknowledges support from the DFG SFB-668 and The Hamburg Centre for Ultrafast Imaging,
 M.~I.~K. acknowledges financial support from ERC (project 338957 FEMTO/NANO) and from NWO via the Spinoza Prize.
Computations have been performed at the NIC, Forschungszentrum J\"ulich, under project HH14.

{}
\end{document}